# Comparison of a Self-Limiting Transformer and a Transformer Type FCL with HTS Elements

Vladimir Sokolovsky, Victor Meerovich, and Istvan Vajda

*Abstract*— A superconducting fault current limiter of the transformer type (inductive FCL) based on magnetic coupling between a superconducting element and a protected circuit has been investigated by many authors for various parameters and performances of a superconducting element. Another design of the device preventing high short-circuit currents is a self-limiting transformer combining the functions of a usual power transformer with the functions of a current limiter. In the presented work we compare the parameters, operation and application of these two devices. The operation of the devices is investigated experimentally on small models fabricated using the same superconducting element. The parameters of the full-scale devices are evaluated. It is shown that the requirements to superconducting elements are practically the same for both devices.

## I. INTRODUCTION

TO reduce fault currents, different fault current limiters (FCLs) were proposed, among them superconducting (SC) devices based on a fast transition from the SC state to the normal state. One design - an inductive limiter based on magnetic coupling between an SC element and a protected circuit - was investigated by many authors for various parameters and performances of a superconducting element [1 and ref. in it]. Typically this device is a two-winding transformer with a primary normal metal winding inserted in series into the circuit to be protected and a SC short-circuited secondary winding, which undertakes the superconducting-normal state transition (S-N transition) under an excess current in the circuit. The secondary coil in an inductive FCL based on high-temperature superconductors (HTS) is frequently performed in the form of a set of HTS rings or cylinders [2, 3]. As an alternative design, multiple-turn SC coils short-circuited by SC switching elements were considered [1, 4]. In the last case only the switching element undertakes the S-N transition under a fault while the secondary coil remains in the SC state. Several high power prototypes of SC FCLs have been built and successfully tested showing feasibility of various proposed concepts for application in power electric systems [1-3]. The inductive design was used as a basis for the development of more complicated power devices combining functions of voltage transformation and the current limitation [4, 5].

In the present paper we consider two devices providing current limitation under faulty conditions: an inductive FCL and a current-limiting transformer (CLT). The design, operation, testing and analysis of applicability of inductive FCLs have been reported in many works [1-4]. In this paper we focus on the properties of a CLT and its basic differences from the FCL.

## II. DESIGN AND OPERATION OF A CLT

The design of a CLT (Fig. 1) was first proposed in [5]. The device constitutes a single-phase transformer with two additional windings $w_3$ and $w_s$ located on the same leg: SC winding $w_s$ is short-circuited by a SC switching element, the second, $w_3$, is counter-connected in series with the main secondary winding $w_2$ so that the voltages induced in the windings are of opposite signs. The numbers of turns in the windings $w_2$ and $w_3$ are chosen so that electromotive forces induced in them are compensated if the SC winding is opened. The designs of the SC winding with a switching element are similar to the design for an inductive FCL. The simplest performance is a set of the hollow cylinders or rings.

As and for a FCL, two regimes of the device operation have to be considered. Under nominal regimes of the protected network, the SC short-circuited winding compensates the magnetic flux in the right leg, $\Phi_2 = 0$ (Fig. 1). Therefore the voltage on winding $w_3$ is zero. Magnetic flux $\Phi_1$ in the central leg is determined by the rated voltage $U_{nom}$ applied to primary winding $w_1$: $\Phi_1 = U_{nom}/\omega w_1$ ($\omega$ is the cyclic frequency). The same flux, $\Phi_3 = \Phi_1$, exists in the left leg. The device operates as a usual single-phase transformer.

Under a fault, the increase of the current in the secondary circuit leads to the increase of the current in the SC winding and it passes into the normal state. The magnetic flux penetrates now in the right leg. As two sections of the secondary windings are counter-connected, the secondary voltage decreases drastically and the fault current is deeply

Manuscript received August 28, 2006. This work was supported in part by the European Commission under Grant "SLIMFORMER".

V. Sokolovsky and V. Meerovich are with the the Department of Physics, Ben-Gurion University of the Negev, P. O. Box 653, Beer-Sheva, 84105 Israel (phone: 972-8647-2458; fax: 972-8647-2903; e-mail: victorm@ bgu.ac.il).

I. Vajda is with the Department of Electric Power Engineering, Budapest University of Technology and Economics, Budapest, H-1111 Hungary (e-mail: vajda@supertech.vgt.bme.hu).

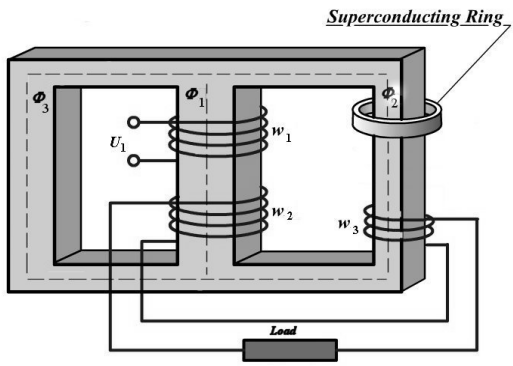

Fig. 1. Schema of a current-limiting transformer.

limited. Under an ideal compensation the secondary current goes to zero. If the resistance of winding $w_s$ is much higher than its inductive reactance the complete compensation is achieved at $w_3 = 2w_2$ [6]. While the FCL action is based on inserting additional impedance in a circuit, the limiting property of a CLT is provided by the compensation of the voltage in the secondary circuit.

### III. EXPERIMENTAL INVESTIGATION

A model has been build using an E-type ferromagnetic core of 150x180x25 mm$^3$ with an air gap of about 1.5 mm. The number of turns in the windings are $w_1 = 200$, $w_2 = 100$, $w_3 = 200$. The SC winding is created by a SC hollow BSCCO cylinder, prepared by Nexans GmbH, with the outer diameter of 70 mm, the wall thickness of 2.5 mm, the height of 25 mm, the critical current density of 970 A/cm$^2$ at 77 K and the critical temperature of 105 K. The model allows us to experimentally investigate the operation of an inductive FCL and a CLT. The FCL is modeled if winding $w_3$ is connected in series into a protected circuit and windings $w_1$ and $w_2$ are opened. The FCL impedances are 1.6 Ω and 17 Ω when the cylinder is in the SC and normal state, respectively. The operation of the FCL model is presented in Fig. 2. The CLT is modeled if all the windings are connected as shown in Fig.1.

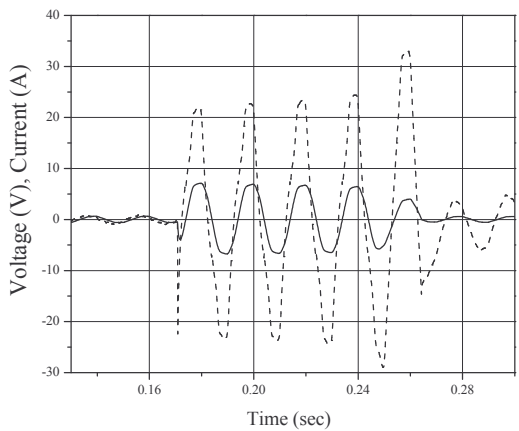

Fig. 2 Oscilloscope traces of current in the protected circuit (solid) and voltage drop across FCL (dashed).

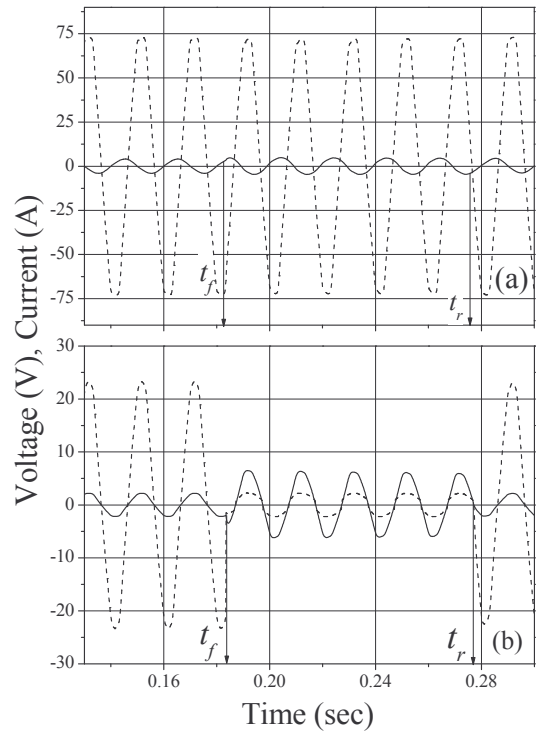

Fig. 3. Operation of the CLT model: oscilloscope traces of the current (solid line) and voltage (dashed line) in the primary (a) and secondary (b) circuits. The secondary voltage is measured across a resistor of about 0.3 Ω. The fault event occurs at $t_f$ and the nominal regime is recovered at $t_r$.

Fig. 3 presents the results of the fault experiment in the circuit with the CLT model.

In both cases of the FCL and CLT applications, there are common features of a current limitation:
- strong limitation of a fault current (non-limited current is higher about 8 times);
- devices limit the first peak of a fault current without appearance of dangerous overvoltages;
- the S-N transition in the cylinder is observed at the current in winding $w_3$ of 3.5 A. This is in well agreement with the theoretical results [6] predicting that the activation current is determined only by the turn number and current in winding $w_3$.

Fig. 2 shows that the FCL impedance increases with time but does not achieve the impedance when the cylinder is in the normal state. It indicates an incomplete S-N transition. Even in this regime the FCL sufficiently limits a fault current. The incomplete S-N transition can also be a cause of moderate limitation (above the nominal current) of a fault current in the secondary circuit of the CLT (Fig. 3b). Another reason can be an incomplete compensation of the electromotive forces induced in windings $w_2$ and $w_3$. The incomplete compensation, relatively large leakage reactance and low magnetizing reactance result in practically unchanged primary current and voltage during a fault (Fig. 3a). In a full-scale device the compensation can be improved by adjustment the turn number in winding $w_3$ with taking into account leakage fluxes and a finite resistance of the SC winding during a fault limitation. At the complete compensation, the primary current would be of the order of the magnetizing current (a percent of the rated current).

IV. CALCULATION OF THE DEVICE PARAMETERS

The requirements and calculation methods for an inductive FCL have been discussed in detail [1]. The parameters of a FCL are determined by the required impedance $x_n$ for current limitation, admissible impedance $x_s$ in the nominal regimes of protected circuit, activation current $i_a$ of the device, and admissible loss value under a fault. Specifically, the reactance of the FCL primary winding $x_n$ (magnetizing reactance) can be varied in a wide range adjusting to the desired level of the fault current limitation. The core diameter, turn number in the primary winding are functions of $x_n$.

The CLT should meet to the requirements to both the transformer and the fault current limiter. Therefore, the magnetizing reactance of the CLT is determined as for a usual transformer from the condition of a small magnetizing current (about 0.5% of the rated current). The core diameter and turn numbers $w_1$ and $w_2$ are determined solely by the power per phase $S$ and the reactance $u_x$ of the transformer with closed secondary winding [7]:

$$D_{core} = K_{tr} \sqrt[4]{\frac{S}{u_x}}; \quad w_1 = K_w \frac{U_{nom}}{D_{core}^2}, \quad (1)$$

where the coefficients $K_{tr}$ and $K_w$ depend on the design features of the windings and magnetic core, isolation gaps, maximum magnetic flux density in the core (for $S$ = 10-30 MVA, $K_{tr} \approx 5 \cdot 10^{-3}$ and $K_w \approx 3 \cdot 10^{-3}$ in units of SI).

The current limitation level in the CLT is determined by the ratio $w_3/w_2$ and can be varied by the change of the turn number $w_3$. However the variation of $w_3$ is also limited [6]: if $w_3 < w_2$ magnetic flux in the right leg under fault becomes higher than in nominal regimes and, therefore, larger diameter $D_{core}$ is needed. On the other hand, the critical current and volume of the SC switching element increase with the turn number $w_3$. At $w_3 = 2w_2$ the secondary current is limited till zero and further increase of the number leads to growth of this current.

The SC switching element for both the devices is calculated similarly. Calculation principles for a FCL can be found in [1]. As differentiated from the FCL, the required critical current of the switching element of the CLT is determined by the multiplication of the turn number in winding $w_3$ and the activation current referred to the secondary winding $w_2$ [6].

Let us evaluate the parameters of full-scale FCL and CLT for the same circuit. Consider the case when an inductive FCL is installed in each phase of a three-phase 35/11 kV, 45 MVA transformer from low-voltage side, i. e. into a circuit with the phase voltage of 6.3 kV and rated current of 2400A. As an alternative, we will consider 3 single-phase 20/6.3 kV, 15 MVA CLT, wye-wye connected. In both cases, the SC windings are performed in the form of hollow BSCCO cylinders which serve also as switching elements. Let the critical current density is $2 \cdot 10^7$ A/m$^2$, the critical temperature is 105 K.

The basic parameters of the FCL and CLT for several different values of the limited current are given in Table 1.

The temperature rise of the cylinder is estimated in adiabatic approximation assuming that homogeneous cylinders fast pass into the normal state and keep this state all the time of limitation. The final temperature of the cylinder after fault clearing (recovery of the nominal load in circuit) is also presented in the table. The estimates of the temperature rise show that not in all the cases the cylinder is heated up to the critical temperature. Therefore the cylinder can be in the resistive state with the resistance lower than the normal state resistance taken in the estimations. To study the operation of the device and the cylinder heating in more details, we performed the simulation of transient processes in a single-phase circuit with the CLT. The simulation was based on the equations of the electrical circuit and on the non-stationary thermal state equation for a homogeneous superconductor:

$$U_{10}\sin(\omega t) = R_{11}i_1 + L_{11}\frac{di_1}{dt} + (L_{12} - L_{13})\frac{di_2}{dt} + L_{1s}\frac{di_s}{dt}$$

$$0 = (L_{12} - L_{13})\frac{di_1}{dt} + R_{22}i_2 + (L_{22} - 2L_{23} + L_{33})\frac{di_2}{dt} + R_{load}i_2 +$$

$$L_{load}\frac{di_2}{dt} + (L_{2s} - L_{3s})\frac{di_s}{dt}$$

$$0 = L_{1s}\frac{di_1}{dt} + (L_{2s} - L_{3s})\frac{di_2}{dt} + L_s\frac{di_s}{dt} + u_s$$

$$C\frac{dT}{dt} = u_s i_s - hSur(T - T_0)$$

(1)

where $U_{10}$ is the primary terminal voltage; $T$ and $T_0$ are the temperatures of the superconductor and the coolant, respectively; $C$ is the heat capacity of the cylinder; $Sur$ is the cooled surface; $h$ is the heat transfer coefficient, $i_i$, $R_{ii}$ and $L_{ii}$ are the current, resistance and inductance of the winding with the turn number $w_i$ ($i$ = 1, 2, 3, $s$), index $s$ related to the superconducting winding; $R_{load}$ and $L_{load}$ is the resistance and self-inductance of the load connected to the secondary winding $w_i$ ($i$ = 2, 3); $L_{ij}$ is the mutual inductance of windings $w_i$ and $w_j$; $u_s$ is the voltage drop across the switching element. The superconductor was approximated by a non-linear voltage-current characteristic including three parts corresponding to the flux creep, flux flow and normal state [1, 9].

The simulation results for the CLT with $w_3 = 2w_2$ =98 are shown in Fig. 4 and the refined values of the temperature rise – in Table 1 in brackets. For the final steady-state temperature after the nominal load recovery, the simulation and estimation results coincide.

V. DISCUSSION

Both devices, FCL and CLT, limit successfully fault currents even in cases when the switching element resistance is lower than the reactance of the SC winding. Limiting effect is achieved also at the incomplete transition of the element into the normal state (Fig. 4a). In this case, the initial stage of the process is characterized by higher limited current and losses in the superconductor. The last explains higher values of the temperature rise obtained by the numerical simulation than those derived from the adiabatic approximation.

TABLE I
PARAMETERS OF CLT AND FCL

| Device | FCL | | CLT | |
|---|---|---|---|---|
| Design | Open core | | Single-phase | |
| Limited current | $2I_{nom}$ | $4I_{nom}$ | $0.92I_{nom}$ | $0.46 I_{nom}$ |
| Inductive reactance of primary winding, Ω | 1.1 | 0.4 | 4000 | 4000 |
| Primary turn number, $w_1$ | 56 | 33 | 155 | 155 |
| Second. turn number, $w_2$ | -- | -- | 49 | 49 |
| Turn number $w_3$ | --- | --- | 49 | 98 |
| Core diameter, m | 0.62 | 0.73 | 0.62 | 0.62 |
| Volume of SC, m$^3$ | 0.082 | 0.025 | 0.065 | 0.032 |
| Normal state resist., Ω | $1.5·10^{-4}$ | $4.8·10^{-4}$ | $6·10^{-4}$ | $1.2·10^{-3}$ |
| Loss power in SC, W | $7.7·10^6$ | $6.8·10^6$ | $7.0·10^6$ | $1.4·10^7$ |
| Temperature rise, K | 28 | 39 | 61 (72) | 15 (32) |
| Temperature at nominal load recovery, K | 35 | 9.7 | 105 | 95 |

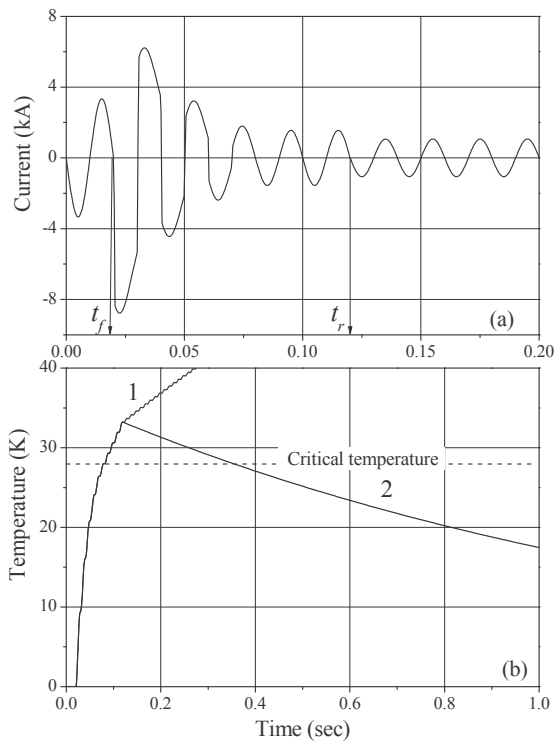

Fig. 4. Simulation results for secondary current (a) and temperature of heating (b): 1- fault and recovery of the nominal load; 2- fault and non-current pause. Times: $t_f$ is the point of fault; $t_r$ is the point of recovery of the nominal load.

As one can see from Table 1, the CLT design based on a SC cylinder leads to the deep current limitation (below $I_{nom}$) for all real relationships between $w_2$ and $w_3$ while the FCL design allows the wide range of limited currents. Recovery of the SC state in the cylinder without non-current pause can be realized only with the FCL (limited current $4I_{nom}$). In other cases the SC state recovery is possible only within a sufficient non-current pause (Fig. 4b).

Note that the deep current limitation is problematic also from point of view of the stability of the SC state. As follows from Table 1, the normal state in the cylinder can exist at currents below the nominal value. Fig. 4 shows that the temperature rise continues when the current is below the rated one. One possible solution of these problems could be the substantial increase of the critical current density leading to the decrease of losses and temperature at recovery of the nominal regime.

One obstacle for building transformers with lower leakage reactance is the necessity to provide an admissible level of fault currents. Unique properties of the CLT- possibility to limit both the transient and steady-state currents - allow one to build transformers with low leakage reactance.